# Anomalous Field-Temperature Phase Diagram of Superconductivity in Sn-Pb Solder

Takumi Murakami[†], Hiroto Arima[†], Yoshikazu Mizuguchi*

*Department of Physics, Tokyo Metropolitan University, 1-1, Minami-osawa, Hachioji 192-0397, Japan*

*Corresponding author: mizugu@tmu.ac.jp

[†]The two authors contributed equally.

**Abstract**

Sn-Pb solders are superconducting materials whose Sn and Pb are perfectly phase-separated. Recently, anomalous magnetic-flux trapping in a Sn45-Pb55 solder has been revealed where the magnetic fluxes are selectively trapped in the Sn regions due to the supercurrents in the surrounding Pb regions. Here, we report on the observation of the anomalous critical field ($H_c$)-temperature ($T$) phase diagram of superconductivity in the Sn45-Pb55 solder. Although the $H_c(T)$ for the Pb regions decreases with increasing field as in normal type-I superconductors and is consistent with the conventional trend with $H_c(0) \sim 800$ Oe, the $H_c(T)$ for the Sn regions exhibits anomalous behaviors. The most noticeable trend was observed in the field-cooled (under 1500 Oe) solder. The $H_c$-$T$ phase diagram for the Sn regions largely varies when the applied external field is reversed, and $T_c$ increases with increasing field amplitude when $H < 0$ Oe is applied. On the basis of the flux trapping and the observed anomalous $H_c$-$T$ phase diagrams, we propose that the field-robust superconductivity in the Sn regions is related to the ferromagnetically-aligned magnetic fluxes (or formation of vortices) in the Sn regions of the Sn45-Pb55 solder.

**Main text**

Superconductivity is a quantum phenomenon where electrical resistivity becomes zero at temperatures ($T$) lower than the transition temperature ($T_c$) of the material. When superconductivity emerges, electrons form Cooper pairs [1] using attractive interactions such as electron-phonon coupling (conventional pairing) or fluctuations of spins or orbitals (unconventional pairing) [2-11]. Due to their exotic superconducting properties, superconductors with unconventional pairing have been extensively studied. Among them, Uranium-based (U-based) superconductors have been drawing much attention because of the observation of spin-triplet superconductivity and extremely high (robust) upper critical fields ($H_{c2}$) [10-17]. In the U-based superconductors, ferromagnetism and superconductivity can coexist, and inhomogeneous superconducting states are observed, for example, in UCoGe [18]. Although the U-based superconductor systems have provided various exotic characteristics, such spin-triplet superconductivity and/or the coexistence of ferromagnetism and superconductivity has been revealed in limited material systems including organic conductors [19,20]. Therefore, researchers have explored different types of superconductors, in which ferromagnetism and bulk superconductivity may coexist [21]. Furthermore, ferromagnet-superconductor hybrid structures have been studied [22-24], and the modification of superconducting states and inhomogeneous superconducting states were observed [24,25]. Therefore, to explore a new platform for studying pure and applied physics on ferromagnetism-related superconductivity, developments of new materials and new concept have been desired.

In this letter, we report on the observation of anomalous critical field ($H_c$)-temperature ($T$) phase diagrams of the superconducting states of a commercial Sn45-Pb55 solder (mass fraction: Sn:Pb = 45:55, atomic ratio: $Sn_{0.58}Pb_{0.42}$) without a flux core. Recently, superconducting and thermal transport properties of the Sn-Pb solders were carefully investigated because of the observation of nonvolatile magneto-thermal switching with a large switching ratio [26]. The phenomenon is caused by magnetic-flux trapping in the Sn-Pb solders [26-28]. The Sn-Pb solders are perfectly-phase-separated materials containing μm-scale Sn and Pb regions, and the magnetic fluxes are trapped by field cooling (FC) under fields greater than $H_c$ of Sn [26]. The temperature dependence of magnetization clearly showed that the field-cooled solder exhibits a ferromagnetic-like behavior below 7.2 K ($T_c$ of the Pb regions). Figure 1 shows the schematic images of the absence/presence of the trapped fluxes after (a) zero-field cooling (ZFC) and (b) FC under 1500 Oe. The fluxes are trapped in the normal-conducting Sn regions at $T < T_c$ (Pb). In this study, we focused on the presence of ferromagnetically-trapped magnetic fluxes in the Sn45-Pb55 solders and explored the possibility of the coexistence of the fluxes and bulk superconductivity in the Sn regions.

The Sn45-Pb55 solder used in this study is a product of TAIYO ELECTRIC IND. CO., LTD. and contains tiny (0.2%wt.) Cu impurity [26]. Specific heat measurements were performed on a Physical Property Measurement System (PPMS Dyna Cool, Quantum Design) by a relaxation method. The weight of the measured sample is 14.5 mg, which was placed on the sample puck with N grease. We measured the temperature dependences of specific heat after ZFC and FC under $H$ = 1500 Oe. For the FC case, data under positive (same direction as the applied field in the FC process; $H > 0$ Oe) and negative (opposite direction as the applied field in the FC process; $H < 0$ Oe) magnetic fields were taken in different sequences after the same FC process.

We measured specific heat of the Sn45-Pb55 solder with a $H$ interval of 100 Oe after flux trapping by FC under 1500 Oe. According to previous works, the maximum quantity of trapped fluxes in Sn-Pb solders is around 500 Oe [26-28]. Therefore, FC under 1500 Oe can produce enough magnetic fluxes in the Sn regions. Figure 2(a) shows selected data of the temperature dependences of $C/T$ under positive fields. Because of Debye temperature lower than 130 K [26], the contributions from phonon are already dominant at the investigated temperature range, but anomalies at $T_c$ of Pb are clearly observed in Fig. 2(a). In addition, the other anomalies are observed at lower temperatures. To analyze the contributions of superconducting transitions in the solder, $[C-C(1000\ \text{Oe})]/T$ data are plotted as a function of $T^2$ in Figs. 2(b); subtracting $C(1000\ \text{Oe})$ data can highlight the evolutions of superconducting states because the solder is in normal states at $H = 1000$ Oe. The data obtained under negative fields are summarized in Figs. 2(c) and 2(d) in the same manner as Figs. 2(a) and 2(b). To discuss the systematic evolution of the transitions with field, all data taken under $H = $ -800-800 Oe with different origin lines are plotted with an arbitrary unit in Fig. 2(e). Under both positive ($H > 0$ Oe) and negative ($H < 0$ Oe), similar trends were observed, but the amplitude of the transition and the $T_c$, particularly for the low-$T$ anomalies, were differed between positive and negative fields. The $T_c$ of Pb is conventionally suppressed for both cases, but the transition temperature of the low-$T$ anomaly exhibits asymmetric evolutions to the positive and negative fields. The transition temperature increases with decreasing fields from positive ($H = 800$ Oe) to negative ($H = $ -700 Oe). The increase in transition temperature with increasing field amplitude of the negative fields ($H < 0$ Oe) implies that the states of the Sn regions are feeling two different fields, inner fields and external fields. It is natural to consider that superconductivity emerges in the Sn regions, and the $T_c$ increased by the compensation of the fields, which results in the $T_c > 3$ K, which is close to $T_c = 3.7$ K for pure Sn. We consider that the transitions observed at low temperatures are originated from the emergence of superconductivity in the Sn regions.

In Fig. 3, the $H_c$-$T$ phase diagrams for $H > 0$ Oe and $H < 0$ Oe are plotted together. The $H_c$-$T$ curve for the Pb region nicely agrees with the dashed line, which describes $H_c(T) = H_c(0)[1-(T/T_c)^2]$, with $H_c(0)$ = 800 Oe (or -800 Oe) of pure Pb. However, $T_c$ of the Sn regions clearly deviates from the dashed line drawn by assuming $H_c(0)$ = 300 Oe (or -300 Oe). For data under $H$ > 0 Oe, quite robust superconductivity to fields was observed at low temperatures. Furthermore, in the phase diagram for $H$ > 0 Oe, $T_c$ increases with increasing fields down to $H$ = -400 Oe, which is close to the absolute value of the trapped field. Therefore, we consider that the increase in $T_c$ with negative fields would be explained by the compensation of the effective fields near the Sn superconducting regions. Noticeably, the unconventional phase diagram is reminiscent of those observed in U-based systems. Also, the phase diagram shows the robustness of $T_c$ to fields for the superconductivity in the anomalous Sn regions. Furthermore, the observation of superconductivity of Sn under $|H|$ > $H_c$ (pure Sn) suggests the emergence of unconventional superconducting states in the Sn regions.

Figures 4(a)-4(d) show the results for the ZFC case. The solder sample was heated to $T$ = 8.5 K and then cooled to $T$ = 1.85 K under $H$ = 0 Oe. Therefore, no flux is initially trapped as shown in Fig. 1(a). Figure 4(b) displays data under selected $H$, and for comparison, all data taken under $H$ = 0-800 Oe with different origin lines are plotted with an arbitrary unit in Fig. 4(c). In addition to clear superconducting transitions of Pb, which are suppressed with increasing field, broad transitions at $T$ < 3.7 K are observed at $H$ = 0-400 Oe. Interestingly, low-temperature anomalies are observed in Figs. 4(b) and 4(c), which is similar to those observed in the FC data. Sharp transitions are observed at $T$ ~ 2.5 K for data under $H$ = 500 and 600 Oe. At $H$ = 700 Oe, the transition would sit at the same temperature, but the anomaly is mixed with that for Pb. Figure 4(d) shows the $H_c$-$T$ phase diagram (ZFC) with $T_c$ of the Pb and Sn regions. The $H_c$-$T$ curve for the Pb region nicely agrees with the theoretical (dashed) line with $H_c(0)$ = 800 Oe of pure Pb. However, that of the Sn regions clearly deviates from the dashed line drawn by assuming $H_c(0)$ = 300 Oe, and the data shows the robustness of $H_c$ to the applied fields. The anomalous superconducting phases in the Sn regions are commonly induced in both FC and ZFC samples while the characteristics of the phase depend on the applied

magnetic field.

Here, we briefly discuss possible causes of the anomalous-superconductivity states observed in the Sn regions of the Sn45-Pb55 solder. When external magnetic fields are removed or reduced after FC (1500 Oe), magnetic fluxes of the Pb regions are expelled due to Meissner states, and the Sn regions accept the fluxes. This situation should be valid because of the conventional $H_c$-$T$ behavior in the Pb regions. Because of the trapped fluxes exceeding $H_c$ of Sn, the Sn regions cannot undergo a transition to normal Meissner states. In such a situation, however, we observe a sharp superconducting transition, and the $T_c$ of the Sn regions is quite robust to external fields. The most possible scenario is the transformation of the trapped fluxes with normal-state Sn to vortex-like states with superconducting Sn. Several studies revealed that type-I superconductors can be type-II, and vortices are produced after impurity, size, or structural origins [29-31]. If the mechanisms of the emergence of superconducting states of the Sn regions are related to the formation of vortices, the anomalous robustness of $H_c$ to fields would be exotic states. Since the origins of the phenomena are related to the ferromagnetically-trapped magnetic fluxes in the Sn-Pb solder, ideally phase-separated superconductors including Sn-Pb solders will provide us with a new platform for studying superconducting states related to magnetism and the possible coexistence of superconductivity and ferromagnetism. To deeply understand the phenomena reported here, further experimental and theoretical investigations on the superconducting and magnetic properties of Sn-Pb solders under magnetic fields are needed.

In conclusion, we investigated low-temperature specific heat of the Sn45-Pb55 solder under various magnetic fields after FC and ZFC. Unexpectedly, phase transitions, which are originated from the Sn regions, are observed in the FC and ZFC samples. Although the $H_c$-$T$ phase diagram for the Pb regions are conventional (consistent with pure Pb), that for the Sn regions largely deviate from the conventional $H_c$-$T$ expected for pure Sn. In the FC (1500 Oe) case, a large quantity of fluxes is trapped in the Sn regions and ferromagnetically aligned. Once the external fields are removed or reduced, the low-temperature transitions were observed up to $H_c$ of the Pb regions. The $H_c$-$T$ phase diagrams for the superconductivity of the Sn regions exhibit unconventional trends, and $T_c$ is quite robust to fields. In addition, with applying negative fields, $T_c$ increases with increasing field amplitude at $H < 0$ Oe. The causes of the phenomena would be related to the ferromagnetically-aligned fluxes (when Sn is normal conducting) and formation of vortices (when Sn is superconducting).

**Acknowledgements**

The authors thank Y. Watanabe, R. Higashinaka, T. D. Matsuda, and O. Miura for supports in experiments and discussion. This work was partly supported by JST-ERATO (JPMJER2201), TMU Research Project for Emergent Future Society, and Tokyo Government Advanced Research (H31-1).

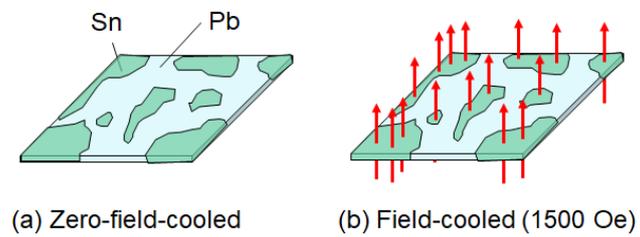

Fig. 1. (Color online) Schematic images of the absence/presence of the trapped fluxes in Sn-Pb solders after (a) zero-field cooling (ZFC) and (b) FC under 1500 Oe.

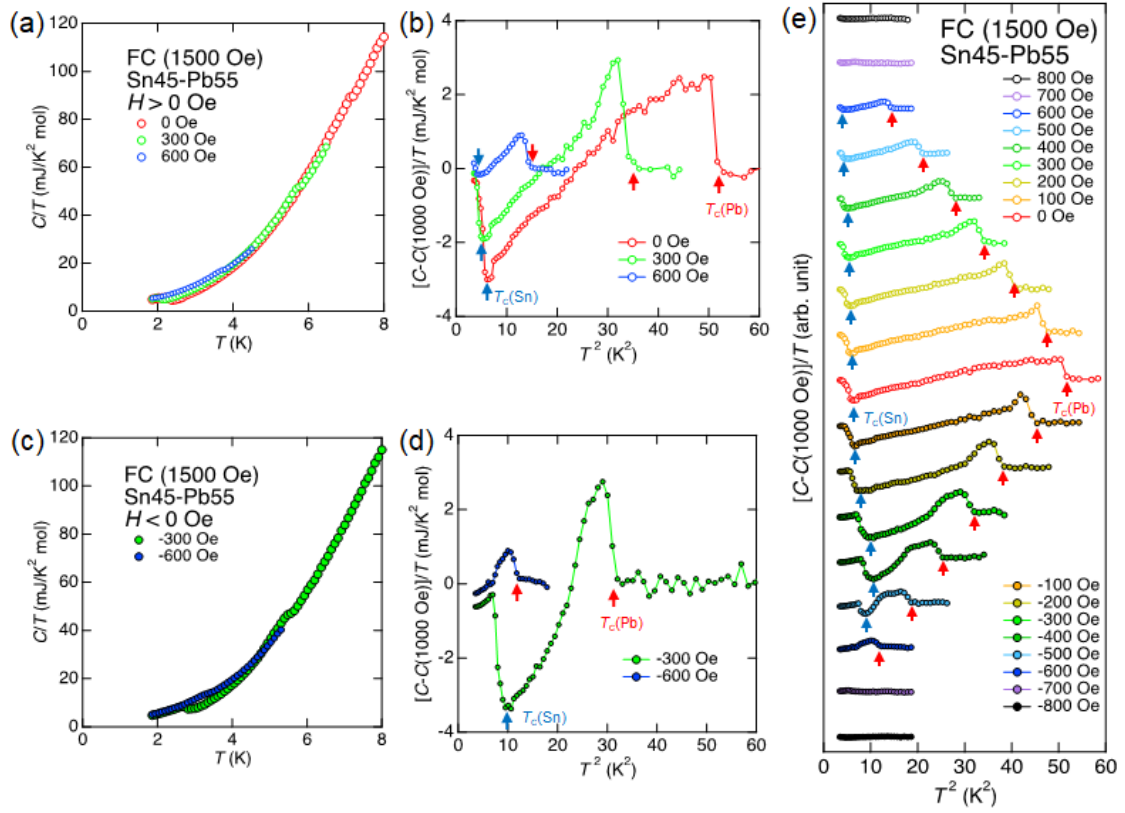

Fig. 2. (Color online) (a,b) Selected $T$ dependences of $C/T$ (after FC under $H = 1500$ Oe) and $T^2$ dependences of $[C-C(1000\text{ Oe})]/T$ where $C(1000\text{ Oe})$ is normal-state data for Sn45-Pb55 under $H = 0$, 300, and 600 Oe. (c,d) Selected $T$ dependences of $C/T$ (after FC under $H = 1500$ Oe) and $T^2$ dependences of $[C-C(1000\text{ Oe})]/T$ under $H = -300$ and $-600$ Oe. (e) $T^2$ dependences of $[C-C(1000\text{ Oe})]/T$ plotted in arb. unit for comparison.

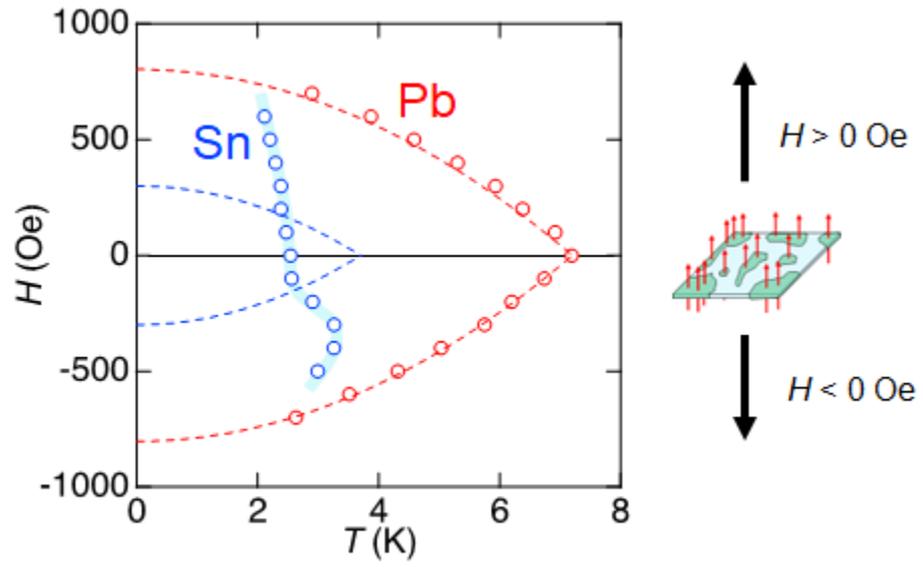

Fig. 3. (Color online) $H_c$-$T$ phase diagram for FC (1500 Oe) solder. The dashed lines are theoretical values estimated using $H_c(0)$ = 300 and 800 Oe (or -300 and -800 Oe) for pure Sn and Pb, respectively. Schematic images for flux-trapping direction and external field direction are shown.

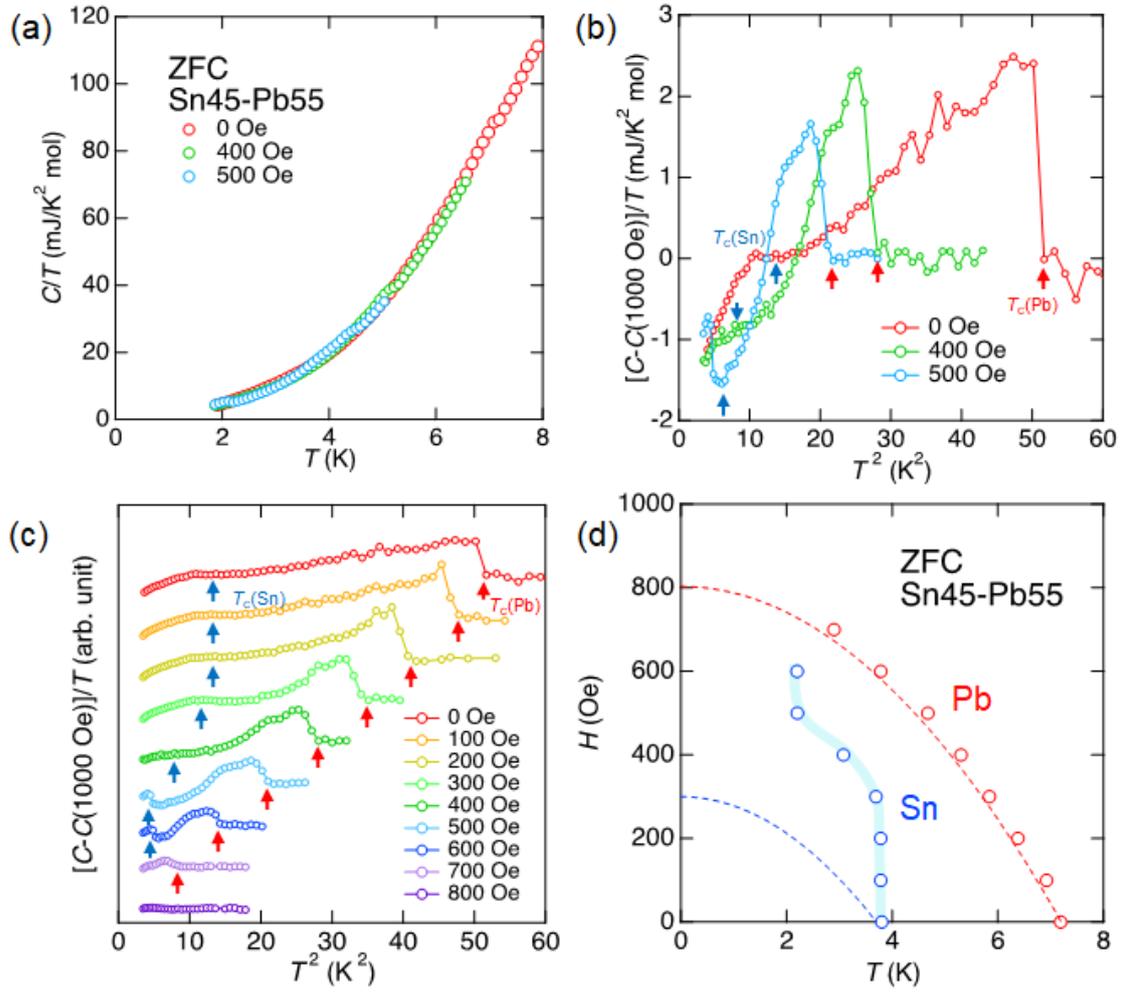

Fig. 4 (Color online) (a) Selected $T$ dependences of $C/T$ (after ZFC) of Sn45-Pb55 at $H$ = 0, 400, and 500 Oe. (b,c) $T^2$ dependences of $[C-C(1000\ \text{Oe})]/T$ where $C(1000\ \text{Oe})$ is normal-state data. (d) $H_c$-$T$ phase diagram for the ZFC case. The dashed lines are theoretical values estimated using $H_c(0)$ = 300 and 800 Oe for pure Sn and Pb, respectively.